\documentclass[12pt, preprint]{aastex}
\usepackage{graphicx}

\begin{document}

\title{The Mass and Radius of the Neutron Star in 4U\,1820$-$30}

\author{Tolga G\"uver, Patricia Wroblewski, Larry Camarota, and
Feryal \"Ozel}

\affil{University of Arizona, Department of Astronomy, 933 N. Cherry 
Ave., Tucson, AZ 85721}

\begin{abstract}
  We report on  the measurement of the mass and  radius of the neutron
  star in the low-mass X-ray binary 4U\,1820$-$30. The analysis of the
  spectroscopic   data  on   multiple   thermonuclear  bursts   yields
  well-constrained  values  for the  apparent  emitting  area and  the
  Eddington flux, both  of which depend in a distinct  way on the mass
  and radius of  the neutron star. The distance to  the source is that
  of  the  globular  cluster   NGC~6624,  where  the  source  resides.
  Combining these measurements,  we uniquely determine the probability
  density over  the stellar mass  and radius. We  find the mass  to be
  $M=1.58  \pm 0.06$~M$_{\sun}$ and  the radius  to be  $R =  9.1 \pm
  0.4$~km.
\end{abstract}


\section{Introduction}

One of the ways to measure the radii and masses of neutron stars is
through a combination of spectroscopic phenomena observed from their
surfaces. In particular, time-resolved spectra of thermonuclear X-ray
bursts provide a measure of the apparent effective area of the neutron
star over a wide range of temperatures.  Furthermore, very luminous
X-ray bursts (called photospheric radius expansion, or PRE, bursts)
cross the local Eddington flux at the neutron star surface and allow
us to obtain a measure of the neutron star mass that is corrected for
general relativistic effects. In the absence of a third spectroscopic
quantity, such as a redshifted absorption line, converting the
observed quantities to masses and radii requires a knowledge of the
distance to the X-ray source.  \"Ozel, G\"uver, \& Psaltis (2009) and
G\"uver et al. (2010) recently measured the masses and radii of the
neutron stars in two low mass X-ray binaries using this approach.

Low mass X-ray binaries located in globular clusters allow for a
convenient application of this method because their distances can be
determined by utilizing the optical and near infrared observations of
the globular clusters (see, e.g., Harris 1996; Valenti, Ferraro, \&
Origlia 2007). Here, we report on the mass and radius of a third
neutron star, the compact object in the binary 4U\,1820$-$30, that
resides in the globular cluster NGC~6624.  4U\,1820$-$30 is an
ultra-compact binary with an orbital period of 11.4 minutes (Stella,
White, \& Priedhorsky 1987). The short period of the system implies a
low-mass, Roche lobe-filling, degenerate dwarf companion with a mass
of about $\sim$ 0.07 M$_{\sun}$ (see, e.g., Stella et al.\ 1987;
Verbunt 1987, Arons \& King 1993; Anderson et al. 1997).

Thermonuclear X-ray bursts from the source were first discovered by
Grindlay et al.\ (1976).  Five type-I X-ray bursts, all showing
evidence of photospheric radius expansion, have been reported from an
analysis of Rossi X-ray Timing Explorer data (Galloway et al. 2008).
4U\,1820$-$30 also exhibited a super-burst during RXTE observations
(Strohmayer \& Brown 2002), most probably resulting from a burning of
a large mass of carbon.  Theoretical modeling of the bursts from
4U\,1820$-$30 agrees well with the observations for a pure He or a
hydrogen-poor companion depending on the assumed value for the
time-averaged accretion rate (Cumming 2003).

In this study, we analyze time resolved X-ray spectra of the
photospheric radius expansion bursts observed from the low mass X-ray
binary 4U\,1820$-$30.  Using the distance measurements to the globular
cluster NGC~6624 and the observed spectral parameters, we determine
the mass of the neutron star in this binary to be $M=1.58 \pm 0.06
M_{\odot}$ and its radius to be $R=9.11 \pm 0.40$~km. Finally, we show
that only smaller value of the two existing distance measurements to
the globular cluster is consistent with the spectroscopic data.

\section{The Distance to NGC 6624}

NGC~6624 is a metal-rich globular cluster with an iron abundance of
$[Fe/H] = -0.63 \pm 0.1$ that is determined from high resolution
infrared spectroscopy with the WFPC2 camera on the Hubble Space
Telescope (Heasley et al.\ 2000; see also Piotto et al. 2002). This
makes its metallicity very close to that of the globular cluster
47~Tuc.  Currently, there are two separate distance measurements to
this globular cluster, performed in optical and near-IR bands.

The first of these measurements is by Kuulkers et al. (2003), based on
the observations of Heasley et al. (2000) in the optical band. Using a
foreground reddening of $E_{B-V} = 0.32 \pm 0.03$, the apparent
magnitude of the horizontal branch $V_{HB} = 16.10 \pm 0.05$, and the
metallicity quoted above, they derive a distance of $7.6 \pm
0.4$~kpc. In this measurement, the absolute calibration of distance is
carried out according to the relation given by Harris (1996).

The second and more recent measurement is reported as part of a
systematic exploration of the near-IR properties of 24 globular
clusters in the galactic bulge (Valenti et al.\ 2007). In their
catalog\footnotemark[1], Valenti et al. (2007) present the photometric
properties, the reddening, and the distances of these clusters.  For
NGC~6624, they find a reddening and a distance of $E_{B-V} = 0.28$ and
$D = 8.4$~kpc, respectively. These results are based on a comparison
of color-magnitude diagrams and the luminosity functions of a cluster
with that of a reference cluster. For metal-rich clusters such as
NGC~6624, 47~Tuc is taken as the reference cluster. The combined error
on the distance modulus is taken to be $\sim$0.15~mag, which includes
the error on the comparison of the color-magnitude diagrams and the
luminosity functions with those of 47~Tuc as well as the uncertainty
of the distance to the reference cluster (Valenti, Private
Communication). This translates to a distance error of 0.6~kpc.

\footnotetext[1]{http://www.bo.astro.it/$^\sim$GC/ir\_archive.}

The two distance measurements are consistent within their formal
uncertainties. In our calculations, we give equal probability to each
and allow for the distances $d=6.8 - 9.6$~kpc, spanning a range that
is 2-$\sigma$ from the central values of each measurement.

\section{Spectral Analysis of X-ray Bursts}

4U\,1820$-$30 was observed with RXTE for a total of 1230 ks until June
2007 (Galloway et al. 2008). However, the majority of the observations
were performed  when the persistent flux  of the source  was above the
critical limit,  beyond which thermonuclear bursts  are not triggered.
Consequently, only  5 thermonuclear X-ray bursts were  detected in the
data set (Zdziarski et al. 2007;  Galloway et al. 2008).  The ratio of
the persistent flux to the peak burst flux during these five bursts is
in the 5\% to 7\% range.

We  extracted time  resolved  2.5$-$25.0~keV X-ray  spectra using  the
ftool  {\it  seextrct}  for  the  science event  mode  data  and  {\it
  saextrct} tool for the science array mode data from all the RXTE/PCA
layers. Science mode observations  provide high count-rate data with a
nominal time resolution of 125$\mu$s  in 64 spectral channels over the
whole energy  range (2$-$60~keV)  of the PCA  detector. We  binned the
X-ray spectra in 26 spectral  channels and over 0.25 seconds for count
rates  above  6000~ct~s$^{-1}$ and  over  0.5  seconds  for count  rates
between 6000$-$3000~ct~s$^{-1}$. Some data gaps during the observations,
however,  necessitated the  extraction of  X-ray spectra  with smaller
exposure times  in a few cases.  We also extracted a  spectrum for the
16~s  intervals prior  to the  bursts and  used it  as  background. We
generated  separate response  matrix files  for each  burst  using the
PCARSP  version  11.7,  HEASOFT  release  6.7,  and  HEASARC's  remote
calibration database and took into  account the offset pointing of the
PCA during  the creation  of the response  matrix files.   This latest
version of  the PCA response  matrix makes the  instrument calibration
self-consistent over  the PCA lifetime  and yields a  normalization of
the  Crab  pulsar  that   is  within  1-$\sigma$  of  the  calibration
measurement of Toor \& Seward (1974) for that source.

To   analyze   the  spectra,   we   used   the  Interactive   Spectral
Interpretation  System  (ISIS), version  1.4.9-55  (Houck \&  Denicola
2000). We  fit each spectra with  a blackbody function  using the {\it
  bbodyrad} model  (as defined  in XSPEC; Arnaud  1996) and  with {\it
  tbabs}   (Wilms,  Allen,   McCray\  2000)   to   model  interstellar
extinction. In  each fit, we included  a systematic error  of 0.5\% as
suggested          by           the          RXTE          calibration
team\footnote[2]{http://www.universe.nasa.gov/xrays/programs/rxte/pca/doc/rmf/pcarmf-11.7/}.

When  fitting  the  burst  spectra,   we  made  use  of  the  detailed
interstellar  extinction measurements towards  4U\,1820$-$30 performed
by Juett, Schulz, \& Chakrabarty  (2004) and Juett et al. (2006) using
Chandra  high  energy transmission  grating  observations. The  column
density values for  O, Ne, and Fe that were  measured in these studies
by  modeling  the  K  and  L  shell  absorption  edges  are  shown  in
Table~\ref{nhtable}.   Note that  the 2-$\sigma$  error bars  given in
Juett et al. (2004, 2006) have been converted to 1-$\sigma$ error bars
in Table~\ref{nhtable} for consistency with  the rest of the paper. We
also converted  the column  density of each  element to  an equivalent
hydrogen column density  using ISM abundances (Wilms et  al. 2000), as
shown  in the last  column.  Given  the statistical  agreement between
these  measurements, we  calculated their  weighted average  and found
$N_{\rm H} = (2.5 \pm  0.3) \times 10^{21}$~cm$^{-2}$, where the error
denotes the  1-$\sigma$ uncertainty.   When fitting burst  spectra, we
fixed the  hydrogen column  density at this  value in the  {\it tbabs}
model.  We note that  converting the reddening, $E(B-V)=0.32 \pm0.03$,
measured by Heasley  et al. (2000) using the  conversion $N_{\rm{H}} =
(6.86 \pm 0.27) \times  10^{21}\times E(B-V)$ (G\"uver \& \"Ozel 2009)
yields  an equivalent  hydrogen  column density  of  $(2.20 \pm  0.22)
\times 10^{21}$~cm$^{-2}$, which is  in agreement with the value given
above.

\begin{figure*}
\centering
   \includegraphics[scale=0.75]{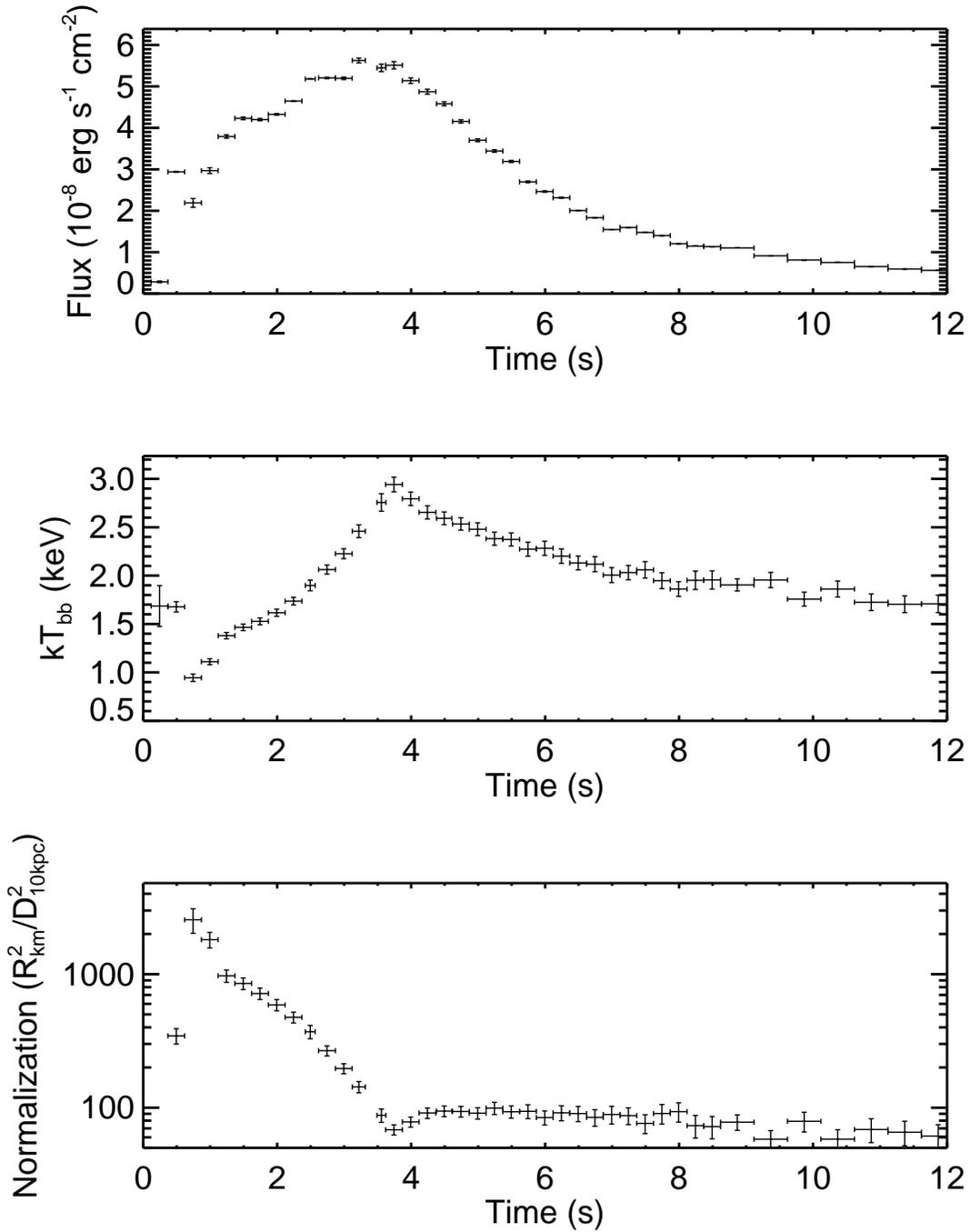}
   \caption{Spectral evolution during the first 12 seconds of an
     example burst of 4U\,1820$-$30 (Obsid : 90027-01-03-05).}
\label{spec_ev_f1820}
\end{figure*}

We found that all five  bursts observed from 4U\,1820$-$30 in the RXTE
archive  were  photospheric  radius  expansion bursts,  following  the
characteristic  evolution  of  temperature,  flux,  and  normalization
associated  with  such   bursts.   In  Figure~\ref{spec_ev_f1820},  we
present  the evolution of  the spectral  parameters during  an example
photospheric  radius  expansion  burst  from  this  source.   We  used
equation ~(3) of  Galloway et al.  (2008) to  calculate the bolometric
fluxes.  All the errors given show the 1-$\sigma$ confidence levels of
our fits.

\begin{figure*}
\centering
   \includegraphics[scale=0.75]{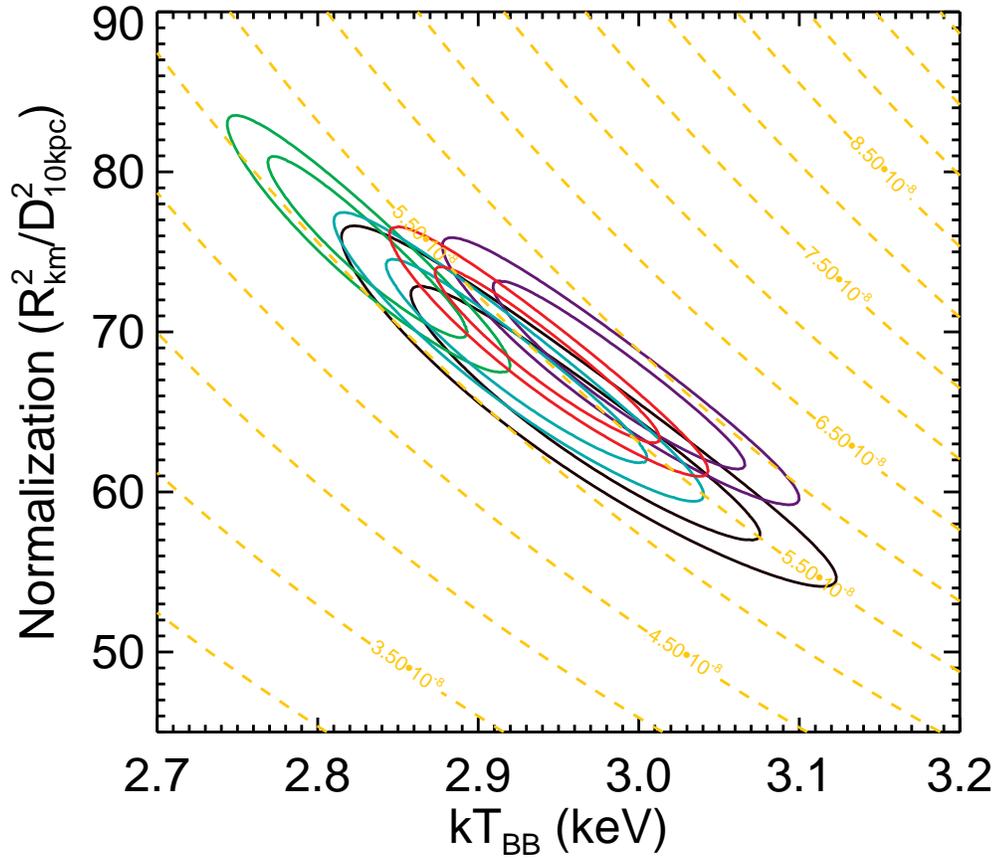}
   \caption{1 and 2-$\sigma$ confidence contours of the normalization
     and blackbody temperature obtained from fitting the five X-ray
     spectra extracted from the touchdown moments of 5 PRE bursts
     observed from 4U\,1820$-$30. The dashed lines show contours of
     constant bolometric flux.}
\label{touchdown1820}
\end{figure*}

Following \"Ozel et al.  (2009) and G\"uver et al.  (2010), we
determined the touchdown flux as the bolometric flux measured when the
normalization of the blackbody obtains its lowest value and the
temperature reaches its highest value.  In Table~\ref{touchdownf} and
Figure~\ref{touchdown1820}, we show the results of the touchdown flux
measurements. Even though the errors in the blackbody normalization
and the spectral temperature are correlated, the touchdown flux
involves a product of these quantities.  This is shown in
Figure~\ref{touchdown1820} as dashed curves corresponding to constant
touchdown flux levels, and can be determined with much higher
precision than either quantity.  All five PRE bursts show touchdown
fluxes that are statistically consistent with each other.  We obtained
the distribution of touchdown fluxes from the probability
distributions over normalization and temperature and determined the
combined $1-$ and 2-$\sigma$ confidence levels. We found an average
touchdown flux of $(5.39 \pm 0.12)\times
10^{-8}$~erg~cm$^{-2}$~s$^{-1}$.

\begin{figure*}
\centering
   \includegraphics[scale=0.75]{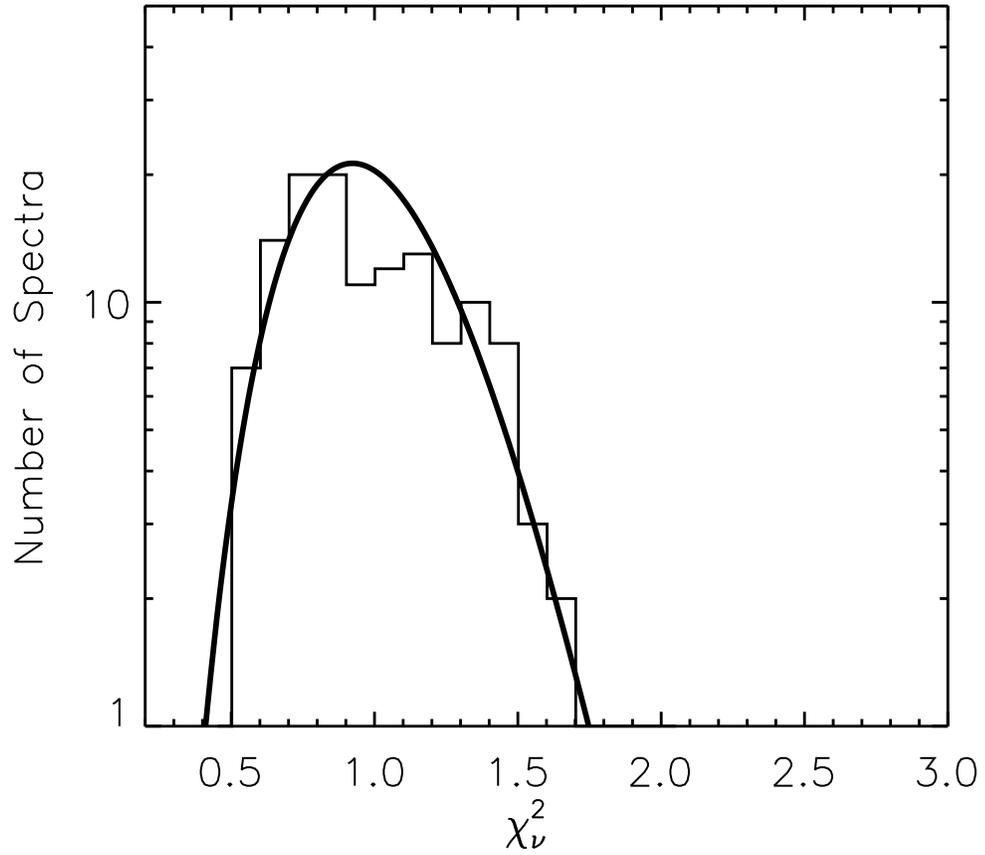}
   \caption{Histogram of the $\chi^{2}_{\nu}$ values obtained from the
     spectral fitting of the X-ray  spectra during the cooling tail of
     all X-ray bursts observed from  the source.  The thick line shows
     the  expected  $\chi^{2}_{\nu}$ distribution  for  26 degrees  of
     freedom.}
\label{chi_hist}
\end{figure*}

\begin{figure*}
\centering
   \includegraphics[scale=0.75]{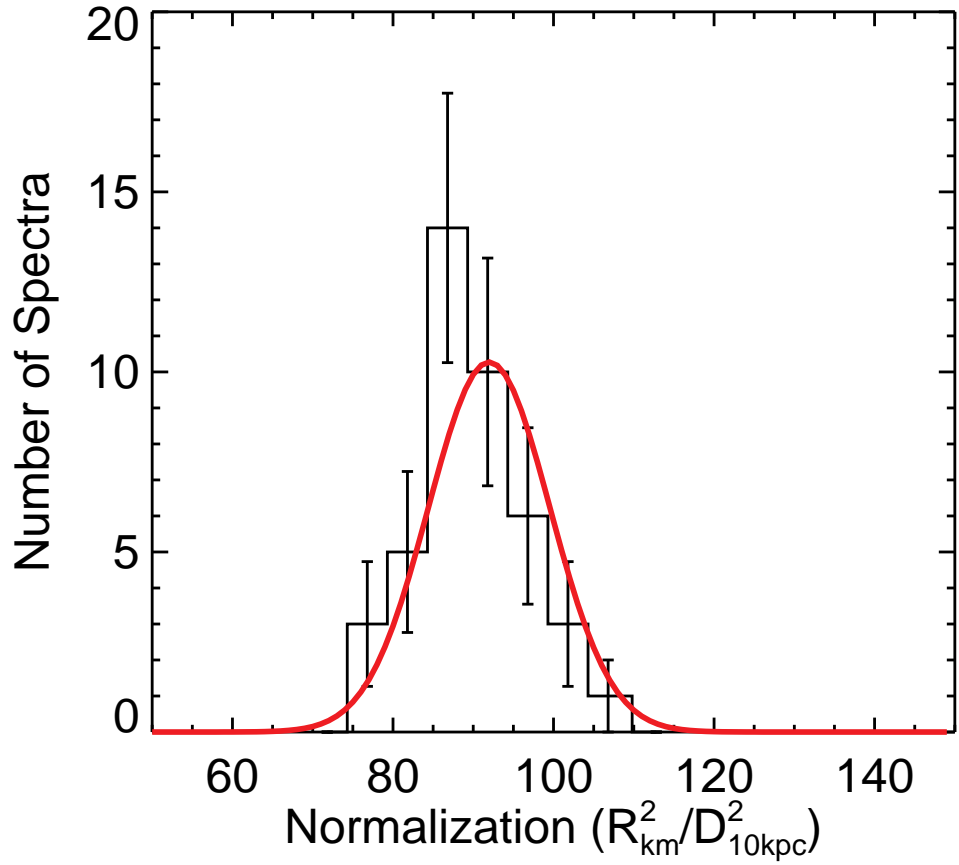}
   \caption{Histogram of  all the normalization  measurements from the
     cooling tails of three X-ray bursts observed from the source. The
     red line  shows the Gaussian  distribution with a sigma  of 11.5,
     which  is similar  to the  average of  the statistical  errors of
     individual measurements.}
\label{norm_hist}
\end{figure*}

\begin{figure*}
\centering
   \includegraphics[scale=0.75]{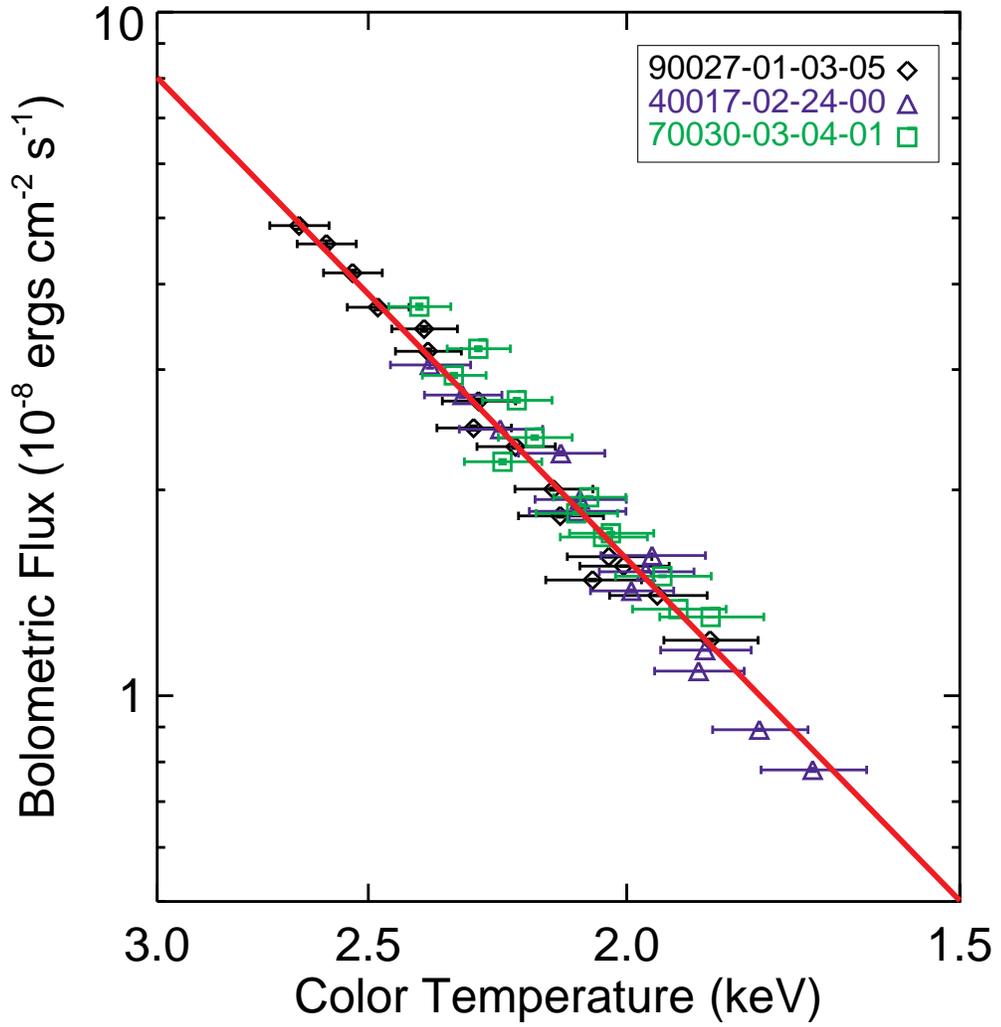}
   \caption{The evolution of the bolometric flux and the blackbody
     temperature during cooling tails of three thermonuclear X-ray
     bursts observed from 4U\,1820$-$30. The bursts follow the $L
     \propto T^{4}$ relation, which is indicated by the solid line.}
\label{flux_kt1820}
\end{figure*}

The second spectral quantity we measure is the normalization of the
blackbody attained during the cooling tails of the thermonuclear
bursts.  The normalization, as defined in the XSPEC model {\it
bbodyrad}, is related to the apparent blackbody radius and the
distance through $A = (R_{\rm app} / D_{10~{\rm kpc}})^2$, where $R$
is the apparent radius of the neutron star and $D$ is the distance in
units of 10~kpc. In Figure~\ref{chi_hist}, we show the distribution of
the $\chi^{2}_{\nu}$ values that we obtained during the cooling tails
of each burst and compare it to the expected distribution for 26
degrees of freedom.  All of the spectral fits follow the expected
distribution and are statistically acceptable.  In order to find the
average normalization for each burst, we took the measurements for the
time intervals $4.25-8$~s after each burst started, when the apparent
radius remained constant while the flux remained high enough for a
reliable determination of spectral parameters.  Constant blackbody
normalizations were clearly observed in three bursts.  In
Figure~\ref{norm_hist}, we show the histogram of all the normalization
values obtained from the blackbody fits and overplot the Gaussian
distribution with a standard deviation equal to the average
statistical error of the individual measurements.  It is evident from
Figure~\ref{norm_hist} that the normalization values follow the
expected normal distribution and are in very good agreement with a
single apparent radius.  Additionally, in Figure~\ref{flux_kt1820}, we
show the evolution of the bolometric flux and the blackbody
temperature during the cooling tails of the three bursts, which also
shows that the evolution of the bolometric flux and the blackbody
temperature values follow the expected $F \propto T^{4}$ relation.

The statistical errors in the spectral parameters increase towards
lower fluxes and this may affect our search for any possible
systematic variations in the normalization. In principle, this effect
can be alleviated by increasing the integration time of each
spectrum. However, given the rapid spectral evolution during X-ray
bursts, using longer exposure time for each X-ray spectrum typically
leads to statistically worse fits because the extracted spectra are
broadened by the variation in the temperature and the source flux.
This may also affect the inferred best fit values.  With that caveat,
we extracted X-ray spectra for longer exposure times (0.5 and 1~s) and
analyzed the re-binned spectra with the same method outlined above.
We then fit the dependence of normalization on temperature with a
linear relation. Even in this case, the data are consistent within
2$\sigma$ with no evolution of the normalization with decreasing
temperature. The best-fit linear model corresponds to a drop in the
normalization from 95.9~(km/10~kpc)$^{2}$ when the temperature was
2.7~keV to 88.7~(km/10~kpc)$^2$ when the temperature dropped to
1.9~keV.  This corresponds to a change of 8\% in the normalization.
In our analysis, we allow for a 30\% systematic error in the
conversion from the normalization to the apparent surface area of the
neutron star by considering color correction factors in the range
1.3$-$1.4 (the apparent area scales as $f_c^4$, see eq.~\ref{ampeq}).
A possible $8\%$ variation in the normalization is within this
systematic uncertainty.

\begin{figure*}
\centering
   \includegraphics[scale=0.75]{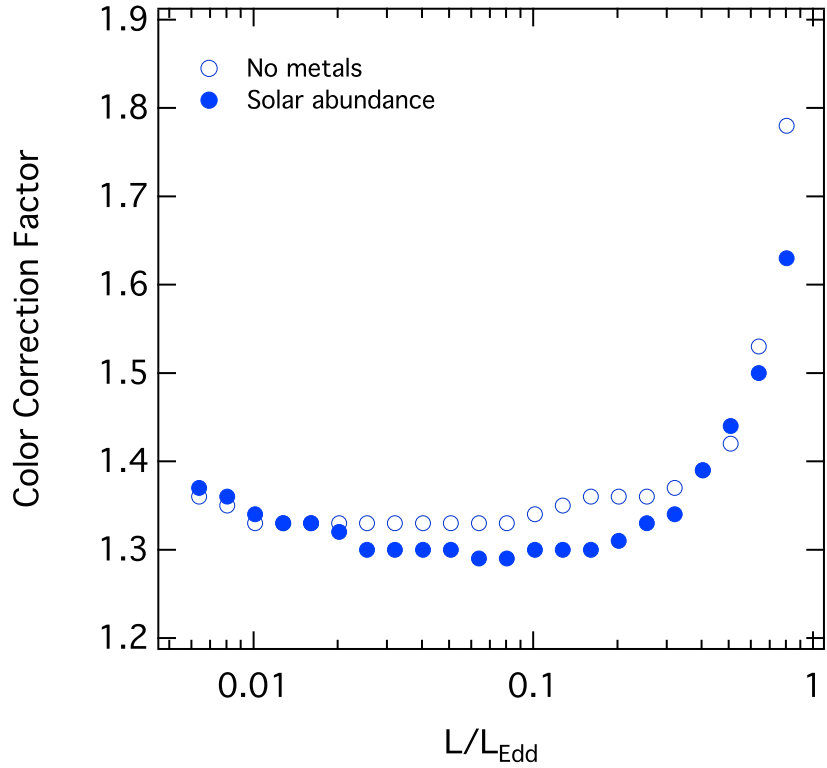}
   \caption{The evolution of the color correction factor
with luminosity, in units of the Eddington luminosity, for atmosphere models with 
zero metallicity (open circles) and with solar metal abundances (filled circles).
These data points correspond to two sequences of models from the calculations of
Madej et al. (2004; zero metallicity) and Majczyna et al. (2005; solar abundances)
that span the widest range of luminosities.}
\label{var_fc}
\end{figure*}

The color correction factor is indeed expected to increase at low flux
levels, and lead to a reduction in the apparent surface area. For
reasonable values of metal abundances in the photosphere, such small
variations in the blackbody normalization at low fluxes are apparent
in the calculations of Madej et al. (2004) and Majczyna et al.  (2005)
(see Figure \ref{var_fc}).

The remaining two bursts had, however, more significant and continuous
fluctuations in the  apparent radii, while one of  the two also showed
an  overall declining trend.   Both the  fluctuations and  the overall
trend cannot be explained by  a simple spectral evolution but indicate
uneven burning  or cooling in  the burst layer.  Therefore,  these two
bursts cannot be used to measure  the radius of the neutron star.  We,
thus, excluded  them from radius measurements.  The  results are shown
in  Table~\ref{touchdownf}. Formally  fitting  the three  measurements
with a constant,  we obtained a blackbody normalization  of $91.98 \pm
1.86$~(km/10~kpc)$^{2}$.

\section{Measurement of the Mass and the Radius of the Neutron Star}

The observed spectral quantities depend on the stellar mass and radius
according to the equations (see, e.g., \"Ozel et al. 2009).

\begin{equation}
  F_{TD} = \frac{GMc}{\kappa_{es}D^2} \left(1-\frac{2GM}{Rc^{2}}\right)^{1/2}
\label{tdeq}
\end{equation}

\begin{equation}
  A = \frac{R^{2}}{D^2f_{c}^{4}} \left(1-\frac{2GM}{Rc^{2}}\right)^{-1},
\label{ampeq}
\end{equation}

\noindent where  G is  the gravitational constant,  c is the  speed of
light,  $\kappa_{es}$  is  the  opacity to  electron  scattering,  and
$f_{c}$ is the color correction factor.

We assign independent probability distribution functions to the
distance P(D)dD, the touchdown flux P(F$_{TD}$)dF$_{TD}$,
normalization P(A)dA, the hydrogen mass fraction P(X)dX, and the color
correction factor P(f$_{c}$)df$_{c}$ that are based on the
measurements, theoretical models, or known priors.  We then calculate
the total probability density over the neutron star mass M and radius
R by integrating the equation

\begin{eqnarray}
&& P(D,X,f_{c},M,R)dDdXdf_{c}dMdR= \frac{1}{2}P(D)P(X)P(f_{c})
P[F_{TD}(M,R,D,X)] \nonumber \\
&& \qquad P[A(M,R,D,f_{c})]J(\frac{F_{TD},A}{M,R})dDdXdf_{c}dMdR\;,
\label{probcalc}
\end{eqnarray}

\noindent over the distance, the  hydrogen mass fraction and the color
correction factor. Here, J(F$_{TD}$,A/M,R) is the Jacobian of the
transformation from the variables (F$_{TD}$, A) to (M,R).

The measurements of the touchdown flux and the normalization were
presented in \S3. Because the different measurements in both cases
agree within their formal statistical errors, we assume a Gaussian
probability distribution for each of these measurements. For the
touchdown flux, we take

\begin{eqnarray}
P(F_{TD})dF_{TD} =
\frac{1}{\sqrt[]{2\pi\sigma^{2}_{F}}}\exp[-\frac{(F_{TD}-F_{0})^{2}}{2\sigma^{2}_{F}}]
\label{tdgauss}
\end{eqnarray}

\noindent  with $F_{0}  = 5.39  \times 10^{-8}$~erg~cm$^{-2}$~s$^{-1}$
and   $\sigma_{F}  =   0.12   \times  10^{-8}$~erg~cm$^{-2}$~s$^{-1}$.
Similarly,  for the  normalization, we  assume a  Gaussian probability
distribution, i,e.,

\begin{eqnarray}
P(A)dA = 
\frac{1}{\sqrt[]{2\pi\sigma^{2}_{A}}}\exp[-\frac{(A-A_{0})^{2}}
{2\sigma^{2}_{A}}]
\label{norgauss}
\end{eqnarray}

\noindent with A$_{0}$ = 91.98~(km/10~kpc)$^2$ and $\sigma_{A}$ = 1.86
(km/10 kpc)$^2$.

Previous studies have yielded two different distance measurements for
the globular cluster NGC~6624, $7.6 \pm 0.4$~kpc from optical
(Kuulkers et al. 2003) and $8.4\pm0.6$~kpc from near-IR observations
(Valenti et al. 2007). In the absence of any further constraint on the
distance to the cluster, we assume a box-car probability distribution,
allowing it to cover the range from 6.8 to 9.6 kpc, i,e.,

\begin{eqnarray}
P(D)dD = \left \{
\begin{array}{lr}
\frac{1}{\Delta D} & {\rm if~} |D-D_{0}| \leq \Delta D/2 \\
0 & {\rm otherwise~}. \\
\end{array} \right.
\label{distanceeq}
\end{eqnarray}

\noindent based on the errors provided by each measurement.

The color correction factor that is obtained from modeling the hot
atmospheres of accreting, bursting neutron stars was discussed in
detail in G\"uver et al. (2010). The calculations show that when the
thermal flux is in the range between $\approx 1\% - 50\%$
sub-Eddington, the color correction factor shows little dependence on
surface gravity or temperature and asymptotes to a well-determined
value (e.g., Madej, Joss, \& Rozanska\ 2004; also see Figure 11 in
G\"uver et al. 2010). Because the color correction is applied to
spectra during the cooling tails of the bursts when the flux is indeed
significantly sub-Eddington, we adopt a color correction factor of
$f_{c} = 1.35 \pm 0.05$ that is appropriate for this regime and
accounts for the range of computed values. We, thus, take a box-car
probability distribution covering the range $1.3-1.4$ so that

\begin{eqnarray}
P(f_{c})d f_{c} = \left \{
\begin{array}{lr}
\frac{1}{\Delta f_{c}} & {\rm if~} |f_{c}-f_{c0}| \leq \Delta f_{c}/2 \\
0 & {\rm otherwise~}. \\
\end{array} \right.
\label{ccorrection}
\end{eqnarray}

\noindent  where $f_{c0}$  =  1.35 and  $\Delta  f_{c}$=0.1 as  stated
above. 

We use the electron scattering opacity $\kappa_{es} = 0.20(1+X)$
cm$^{2}$ g$^{-1}$, which depends on the hydrogen mass fraction $X$.
There is compelling evidence that the accreted material in
4U\,1820$-$30 is either pure He or hydrogen poor (Nelson, Rappaport,
\& Joss 1986). We, therefore, take the hydrogen mass fraction X to be
0 in this case. Note that allowing the hydrogen mass fraction to vary
between $X=0.0-0.3$ does not affect the final mass-radius contours for
this particular source because there are no consistent (M, R)
solutions for the larger X values.

The probability distribution over the neutron star mass and radius can
then  be obtained  by  inserting each  probability distributions  into
equation~(\ref{probcalc})  and integrating over  the distance  and the
hydrogen  mass fraction.  Figure~\ref{massradius} shows  the  $1-$ and
2-$\sigma$  confidence contours  for the  mass and  the radius  of the
neutron star in 4U\,1820$-$30.

\section{Discussion}

\begin{figure*}
\centering
   \includegraphics[scale=0.75]{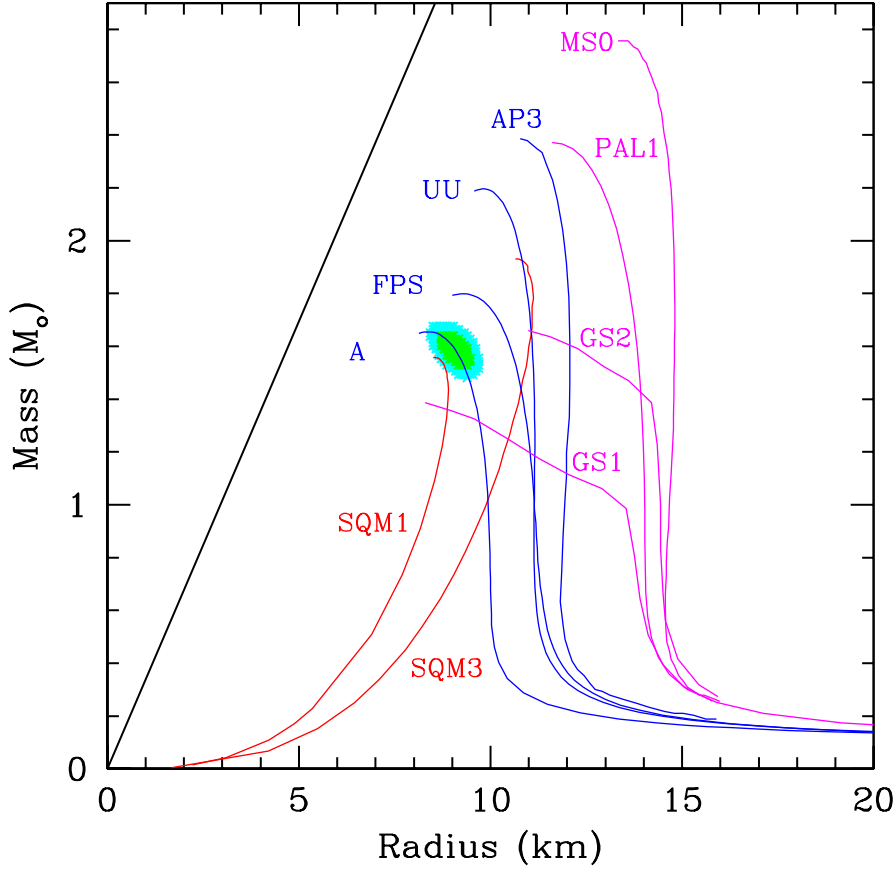} \caption{1 and 2-$\sigma$
   contours for the mass and radius of the neutron star in
   4U\,1820$-$30 are shown together with the predicted mass-radius
   relations for a number of equation of states of neutron star
   matter. The representative mass-radius relations for a select
   number of equations of state include multi-nucleonic ones (A, FPS,
   UU, AP3), equations of state with condensates (GS1-2), strange
   stars (SQM1, SQM3), and meson-exchange models (MS0). The black line
   indicates the black hole event horizon. The descriptions of the
   various equations of state and the corresponding labels can be
   found in Lattimer \& Prakash (2001) and Cook, Shapiro \& Teukolsky
   (1994).}
\label{massradius}
\end{figure*}

We used time-resolved X-ray spectroscopy of the thermonuclear bursts
exhibited by the ultra-compact X-ray binary 4U\,1820$-$30, in
conjunction with the distance measurement to its host globular cluster
NGC~6624, to obtain a measurement of the mass and radius of its
neutron star. We present the resulting 1 and 2-$\sigma$ confidence
contours of the 2-dimensional probability density P(M,R) in
Figure~\ref{massradius}.  The peak of the distribution and the
projected 1-$\sigma$ errors correspond to a mass of M$=1.58 \pm
0.06~$M$_{\sun}$ and a radius of R$= 9.11 \pm 0.4$~km.

Given the relatively large uncertainty in the source distance, the
small uncertainties in the measured mass and radius call for an
elucidation. The probability density over mass and radius is found by
Bayesian analysis, which assigns a probability to each (M,R) pair
based on the likelihood that the measured touchdown flux and the
apparent emitting area can be simultaneously reproduced by that mass
and radius pair, for a given distance.  In the case of 4U\,1820$-$30,
the likelihood drops rapidly towards larger source distances, making
the touchdown flux and the apparent emitting area practically
inconsistent with each other, for any (M,R) pair. Thus, the smaller
distance to the globular cluster is a posteriori favored by the
spectroscopic data.

A mass measurement for the neutron star in 4U\,1820$-$30 was reported
by Zhang et al.\ (1998) (see also Kaaret et al.\ 1999 and Bloser et
al.\ 2000) based on the measurement of the frequencies of kHz QPOs
from that source. In these studies, an apparent flattening of the
dependence of the upper kHz QPO frequency on X-ray countrate was
interpreted as evidence for the accretion disk being truncated at the
radius of the innermost stable circular orbit. The frequency of the
kHz QPO at that instant was equal to $\sim$1060~Hz, which, if
interpreted as a Keplerian frequency at the inner edge of the
accretion disk, resulted in a mass for the neutron star of $\simeq 2.2
M_\odot$.  The interpretation of Zhang et al.\ (1998) has been
questioned later by Mendez et al.\ (1999), who showed that the X-ray
countrate is not a good indicator of mass accretion rate onto the
neutron star.  The highest observed QPO frequency from 4U\,1820$-$30
can, therefore, only be used to place an upper bound on the mass of
the neutron star of $\simeq 2.2 M_\odot$ (Miller, Lamb, \& Psaltis
1998), which is consistent with our mass measurement.

More recently, Cackett et al.\ (2008) reported a measurement of the
inner radius of the optically thick region of the accretion disk in
4U\,1820$-$30 based on the modeling of the relativistically broadened
iron line in the X-ray spectrum of the source observed with Suzaku.
They estimated the inner radius of the disk to be $R_{\rm in}/M=
6.7^{+1.4}_{-0.7}$, where all quantities are expressed in geometric
units. Our best-fit value for the compactness of the neutron star is
$R/M \simeq 3.9$, which is smaller than the value reported by Cackett
et al.\ (2008) for the accretion disk, as expected.  The gravitational
redshift corresponding to this stellar compactness is $z
\simeq 0.43$.

The mass and radius reported here are also consistent with the broad
constraints obtained on neutron star properties by observations of
other low-mass X-ray binaries in quiescence (Rutledge et al. 2002;
Heinke et al.  2006; Webb \& Barret 2007). Comparing with the two
measurements of neutron star masses and radii we reported earlier
using the present method (\"Ozel et al.\ 2009; G\"uver et al.\ 2010),
we note that the neutron star in 4U\,1820$-$30 has the smallest mass
of the three (although the three span a relatively narrow range),
while its radius is comparable to those of the other two neutron
stars. The small differences in the masses may result from the unique
accretion histories of the sources and are anticipated. All three, on
the other hand, are systematically more massive than double radio
pulsars, likely due to their long-lived accretion phases.

We have shown in earlier work that a wide range of ultradense matter
equations of state can be well represented by the pressures specified
at three fiducial densities above the nuclear saturation density. The
pressures at these three densities, in turn, can be measured if a
minimum of three well-constrained neutron star masses and radii are
available (\"Ozel \& Psaltis 2009).  The mass and radius of
4U\,1820$-$30 represents the third such measurement and, in principle,
allows for an inversion from masses and radii to the equation of state
parameters.  The resulting implications for the ultradense matter
equation of state will be explored in a forthcoming paper.

\acknowledgments  We are  grateful  to Dimitrios  Psaltis for  various
discussions.   We   thank  the  anonymous  referee   for  very  useful
suggestions. We thank Elena  Valenti, Sergio Ortolani, and Bill Harris
for  useful  discussions  on  the distance  measurements  to  globular
clusters.  F.\"O.  thanks the  Institute for Theory and Computation at
Harvard  University,   where  this  work  was   completed,  for  their
hospitality.   This   work  was  supported   in  part  by   NSF  grant
AST~07-08640.

\begin{table*}
\centering
\caption{Column density in various elements measured by Juett et al. (2004, 2006) 
  with Chandra HETG observations towards 4U\,1820$-$30.}
\begin{tabular}{ccc}
\hline\hline
Element & $N_{X}$ & Equivalent $N_{\rm H}$ \\
 \hline
 O   & $(1.3 \pm 0.1) \times 10^{18}$ & $(2.7 \pm 0.2) \times 10^{21}$   \\
 Ne  & $(3.3 \pm 0.6) \times 10^{17}$ & $(3.8 \pm 0.7) \times 10^{21}$   \\ 
 Fe  & $(5.1 \pm 0.9) \times 10^{16}$ & $(1.9 \pm 0.4) \times 10^{21}$   \\ 
\hline
\label{nhtable}
\end{tabular}
\end{table*}

\begin{table*}
\centering
\caption{Touchdown flux and normalization measurements for the 
  PRE bursts observed from  4U\,1820$-$30.}
\begin{tabular}{cccc}
\hline\hline
Obs. ID & MJD & Touchdown Flux  & Normalization \\
 & &  (10$^{-8}$ erg~cm$^{-2}$~s$^{-1}$) & (R$_{km}$/D$_{10kpc}$)$^{2}$ \\  
\hline
   20075-01-05-00 & 50570.73110 & 5.36 $\pm$0.27 & --\\
   40017-01-24-00 & 52794.73813 & 5.75 $\pm$0.20 & 88.86 $\pm$3.96 \\
   70030-03-04-01 & 52802.07557 & 5.19 $\pm$0.15 & 96.68 $\pm$3.39\\
   70030-03-05-01 & 52805.89566 & 5.34 $\pm$0.20 & --\\
   90027-01-03-05 & 53277.43856 & 5.51 $\pm$0.17 & 90.40 $\pm$2.69\\
\hline
\label{touchdownf}
\end{tabular}
\end{table*}

\end{document}